\documentclass{WileyMSP-template}
\usepackage[superscript]{cite}
\usepackage{wrapfig}
\usepackage{upgreek}
\usepackage{amsmath}

\begin{document}

\pagestyle{fancy}

\title{Microfluidic gratings for X-ray Phase Contrast Imaging}

\maketitle

\author{Alessandro Rossi$^1$}
\author{Francesco Coccimiglio$^2$}
\author{Antonio Ferraro$^3$}
\author{Tiziana Ritacco$^4$}
\author{Alberto Astolfo$^5$} \\
\author{Michele Giocondo$^{3}$}
\author{Vincenzo Formoso$^{2,3}$}
\author{Raffaele Giuseppe Agostino$^{2,3}$}
\author{Francesco Iacoviello$^{6}$} \\
\author{Ioannis Papakonstantinou$^1$}
\author{Alessandro Olivo$^5$*}

\begin{affiliations}
Dr. Alessandro Rossi \\
$^1$ Photonic Innovations Lab, Department of Electronic \& Electrical Engineering, University College London, Torrington Place, London, WC1E 7JE UK \\
Francesco Coccimiglio \\
$^2$ Department of Physics and STAR Research Infrastructure, University of Calabria, Via Tito Flavio, Rende, 87036, CS, Italy \\
Dr. Antonio Ferraro \\
$^3$ Consiglio Nazionale delle Ricerche - Istituto di Nanotecnologia CNR-NANOTEC, Rende, 87036, Italy \\
Dr. Tiziana Ritacco \\
$^4$ Consiglio Nazionale delle Ricerche - Istituto di Microelettronica e Microsistemi, Rome, 00133, Italy \\
Dr. Alberto Astolfo \\
$^5$ Department of Medical Physics and Biomedical Engineering, University College London, Gower Street, London, WC1E 6BT UK \\
Dr. Michele Giocondo \\
$^3$ Consiglio Nazionale delle Ricerche - Istituto di Nanotecnologia CNR-NANOTEC, Rende, 87036, Italy \\
Prof. Vincenzo Formoso \\
$^2$ Department of Physics and STAR Research Infrastructure, University of Calabria, Via Tito Flavio, Rende, 87036, CS, Italy \\
$^3$ Consiglio Nazionale delle Ricerche - Istituto di Nanotecnologia CNR-NANOTEC, Rende, 87036, Italy \\
Prof. Raffaele Giuseppe Agostino
$^2$ Department of Physics and STAR Research Infrastructure, University of Calabria, Via Tito Flavio, Rende, 87036, CS, Italy \\
$^3$ Consiglio Nazionale delle Ricerche - Istituto di Nanotecnologia CNR-NANOTEC, Rende, 87036, Italy \\
Dr. Francesco Iacoviello \\
$^6$ University College London, Electrochemical Innovation Lab, Department of Chemical Engineering, London, WC1E 7JE, UK \\
Prof. Ioannis Papakonstantinou \\
$^1$ Photonic Innovations Lab, Department of Electronic \& Electrical Engineering, University College London, Torrington Place, London, WC1E 7JE UK \\
Prof. Alessandro Olivo \\
$^5$ Department of Medical Physics and Biomedical Engineering, University College London, Gower Street, London, WC1E 6BT UK \\
a.olivo@ucl.ac.uk
\end{affiliations}

\keywords{Microfluidics, X-ray Phase Contrast Imaging, gratings, X-ray masks, Beam Tracking}

\begin{abstract}
\begin{justify}
Fabrication of X-ray gratings has surged in the last two decades thanks to their vast employment in X-ray Phase Contrast Imaging, an imaging technique able to boost X-ray sensitivity to detect otherwise invisible details. These high aspect ratio devices are usually fabricated by complex, costly, multi-step processes that limit their size and volume scaling. These steps commonly involve UV or X-ray lithography, semiconductor selective etching and high-Z metal plating, usually Au, which require expensive tools and materials. Here we present a proof-of-concept fabrication via soft lithography and Hg infusion of microfluidic X-ray absorption gratings and their performance in biomedical imaging. Such fabrication technique requires fewer, less expensive, and more scalable processes using alternative and more sustainable materials, while showing comparable visibility with their conventional Au-based, solid equivalent. Our results constitute a promising shift in X-ray optics fabrication that could significantly lower barriers to commercialization and accelerate the practical deployment of X-ray Phase Contrast Imaging.
\end{justify}
\end{abstract}

\section{Introduction}
\label{sec:intro}
\begin{justify}
\indent X-ray imaging (XRI) has profoundly impacted modern life over the past century, with applications ranging from medical diagnostics and non-destructive industrial testing to ubiquitous use in scientific research \cite{rontgen1896new, bukreeva2016virtual, villarraga2019x}.  As it is conventionally based on attenuation contrast, thus detecting details with different attenuation coefficient than the background, XRI falls short in detecting low-Z materials (e.g., organic structures). Thus X-ray Phase Contrast imaging (XPCI) has attracted significant attention in the last three decades as an alternative imaging method due to its higher sensitivity to low-Z features \cite{lewis2004medical}. \\ 
\indent In the X-ray regime, most elements' refractive index $n = \delta + i \beta$ is characterized by a real part $\delta$ exceeding the imaginary part $\beta$ by orders of magnitude. This means that higher image contrast can in principle be obtained from the detection of an X-ray beam's refraction induced by an object rather than its attenuation. By exploiting this mechanism, XPCI enhances the sensitivity of conventional XRI, enabling the visualization of otherwise invisible structures such as soft tissues, early-stage tumors, pulmonary emphysema, microfractures in composite materials, and explosives in security scans \cite{castelli2011mammography, willer2021x, shoukroun2020enhanced, partridge2022enhanced}. \\
\indent Despite the significant improvement in contrast, XPCI is limited by its requirement for coherent beams typically not attainable with the X-ray sources used in industry or hospitals. To circumvent this requirement, a series of imaging techniques have been devised based on the implementation of high-aspect ratio (AR = lamellae's height/width $>$ 10) gratings in the imaging set-up.
The high-AR gratings can be optimized to either selectively mask or impart a phase shift to the X-ray beam by tuning the thicknesses and widths of their transparent lamellae and highly absorbing/phase shifting septa, being referred to as absorption grating and phase grating, respectively. Regardless, they both require AR $>$ 10 due to the the incoming X-ray photons' high penetration depth, and very fine lateral dimensions (1 - 100 $\upmu$m) due to the small refraction angles induced at these energies (a few $\upmu$rad). Such extreme critical dimensions significantly hinder their fabrication, which was enabled only in the last two decades by a series of technological improvements and novel fabrication processes, which ultimately allowed the translation of XPCI to laboratory and clinical set-ups \cite{david2007fabrication, pfeiffer2006phase}.   \\ 
\indent From the first hard X-ray gratings fabricated for XPCI applications, characterized by AR $\sim$ 6 \cite{david2007fabrication}, a lot of effort has been made to increase the attainable AR up to $\sim$ 40 \cite{miao2014fabrication, michalska2023fabrication, josell2021bottom}, thickness up to $\sim$ 500 $\upmu$m \cite{rossi2025fabrication}, and to reduce their periods down to sub-$\upmu$m range to accommodate the requirements of different phase-based imaging approaches \cite{miao2014fabrication, michalska2023fabrication}. Nevertheless, the fabrication paradigm for higher-AR gratings has not moved away from the lithography-etching-plating pipeline, with the exception of LIGA i.e.,  lithography, plating, molding to replicate, which however requires highly coherent and brilliant synchrotron radiation \cite{pereira2025quantifying, noda2008fabrication}. In both pipelines lithography represents the bottleneck for the attainable period and aperture, as the photoresist (PR) patterning is limited by optical constraints when PRs are exposed to UV e.g., wavelength, numerical aperture, PR optical properties. E-beam lithography and Laser Interference Lithography have been used to reduce the attainable periods to the sub-$\upmu$m range, at cost of significantly longer processing time for the former and smaller exposure area for the latter \cite{miao2014fabrication, michalska2023fabrication}. As higher ARs are preferred, PRs are often employed as etching masks for the underlying substrate (mostly Si) \cite{rossi2025fabrication, david2007fabrication, josell2020pushing, josell2021bottom, romano2020high}. The most popular methods for deep etching of Si are Deep Reactive Ion Etching (DRIE) and Metal Assisted Chemical Etching (MACE), a dry, plasma-induced and a wet method, respectively. The former can reach high etching depths thanks to the multi-cycle Bosch process and the careful optimization of passivation and etching steps \cite{laermer2020deep}. The latter instead exploits Si chemical oxidation and dissolution in solutions comprising an oxidant, usually H$_2$O$_2$, and HF as an etchant, which constitutes a significant health hazard \cite{li2000metal}. This reaction chain is usually very slow though it can be catalyzed by patterning the exposed Si substrate with noble metals such as Au and Pt, driving the reaction downward \cite{huang2011metal}. Finally, Au is the preferred material as beam absorber/shifter thanks to its high density and refractive index combined with its high reduction potential which makes it relatively easy to plate onto a conductive substrate \cite{romankiw1997plating}. In the X-ray grating context, the substrate is either an etched semiconductor (i.e., Si) or developed PR if a LIGA approach is followed. They both do not favor Au plating due to their high resistivity, therefore requiring an additional step of deposition of a seed layer. Atomic Layer Deposition (ALD) represents the gold standard for conformal seed layer deposition \cite{miao2014fabrication, vila2018towards}, although physical sputtering is sometimes preferred due to its lower cost and overall duration \cite{michalska2023fabrication, rossi2025fabrication}. In MACE and LIGA, a bottom-up growth can be obtained using the metal catalyst or a preliminarily deposited seed layer, respectively \cite{romano2020high, noda2008fabrication}.   \\
\indent Fabrication costs increase and success yield decrease with the number of processes required. Expensive materials (e.g., Au and other noble metals) and sophisticated equipment like DRIE, PVD and ALD tools increase costs too, which would also grow proportionally with grating size, as these processes would take longer and/or require larger and more expensive equipment. The high costs and fabrication hindrances are nowadays one of the main obstacles to the widespread use of XPCI, so that low-cost modulators such as abrasive papers have been already tested, unfortunately underperforming compared to grating-based techniques \cite{kashyap2016experimental, wang2016synchrotron}. Here, we propose a paradigm shift in grating fabrication, shifting the focus on using new processes and materials rather than the optimization of existing ones. A new grating design is presented, employing microfluidic polydimethylsiloxane (PDMS) channels patterned via soft lithography as a substrate and filling the channels with highly absorbing fluid Hg. Hg has comparable atomic number and mass density as Au, making it a viable alternative as absorbing material in the channels (here having the same role as the septa in traditional masks) with the benefit of being flown in the PDMS microfluidic channels via simple pumping. The use of soft lithography for the fabrication of PDMS channels, despite currently reaching lower ARs in this proof-of-concept work, has a series of benefits over the conventional Si lithography and etching: it is more easily scalable and recyclable, it is inexpensive as it does not require intensive use of specialized equipment and allows for multiple replication of each individual mold \cite{de2023facile, xia1998soft}. It also potentially enables the fabrication of flexible and curved gratings, which is currently limited by the Si and Au stiffness to large curvature radii \cite{josell2024bottom}. Flexible gratings would accommodate conical beams from conventional X-ray sources, therefore extending the imaging field of view (FOV) with resulting extended applicability in industrial and clinical set-ups. To this date, larger FOV ($\sim$ 38.5 cm x 2.5 cm) in the first prototypes of human-scale XPCI-CT scans have been reached by grating tiling instead thus suffering from regions of affected contrast in the areas between gratings \cite{schroter2017large}. \\
\indent In this work, we present the fabrication and experimental validation of such microfluidic gratings, which are successfully employed in a Beam Tracking (BT) XPCI setup demonstrating robust performance and yielding high-quality, quantitatively accurate phase-contrast images. The choice of a BT approach, where a single absorption grating is used to selectively mask the X-ray beam, allowed us to: i) individually test our gratings performance (which in the context of refractive imaging techniques such as BT are usually referred to as `mask', a term that will also be used in this work when appropriate), ii) efficiently image biological samples obtaining higher contrast in phase-based images compared to attenuation-based, while iii) simultaneously relaxing the coherence requirements by using a 70 $\upmu$m focal spot, uncollimated and fully polychromatic x-ray source. This alternative X-ray grating fabrication protocol could be instrumental for translating XPCI into real-world applications by reducing its entry cost for clinics and hospitals. As XRI remains the most used diagnostic imaging techniques, unlocking XPCI's potential could have an impact on millions of lives worldwide \cite{united2010unscear}.
\end{justify}
\section{Results}
\begin{justify}
A scheme of the fabrication process used in this work is shown in \textbf{Figure \ref{fig:fabscheme}}.  The fabrication consisted in a common soft lithography protocol (Figure \ref{fig:fabscheme}a-d) using a SU-8 patterned mold transferred on a PDMS covalently bonded to a glass slide, followed by Hg pumping inside the obtained channels via a peristaltic pump (Figure \ref{fig:fabscheme}e). The design chosen for the microfluidic channel is schematized in Figure \ref{fig:fabscheme}f: a serpentine with two infusion/extraction channels to pump in the Hg and pump out the excess, respectively. The microfluidic chip was finally employed for imaging exploiting the Hg channels beam splitting properties. The fabrication processes and the materials used are discussed in more details in the Experimental Section. 

\begin{figure}
    \centering
    \includegraphics[width=.75\linewidth]{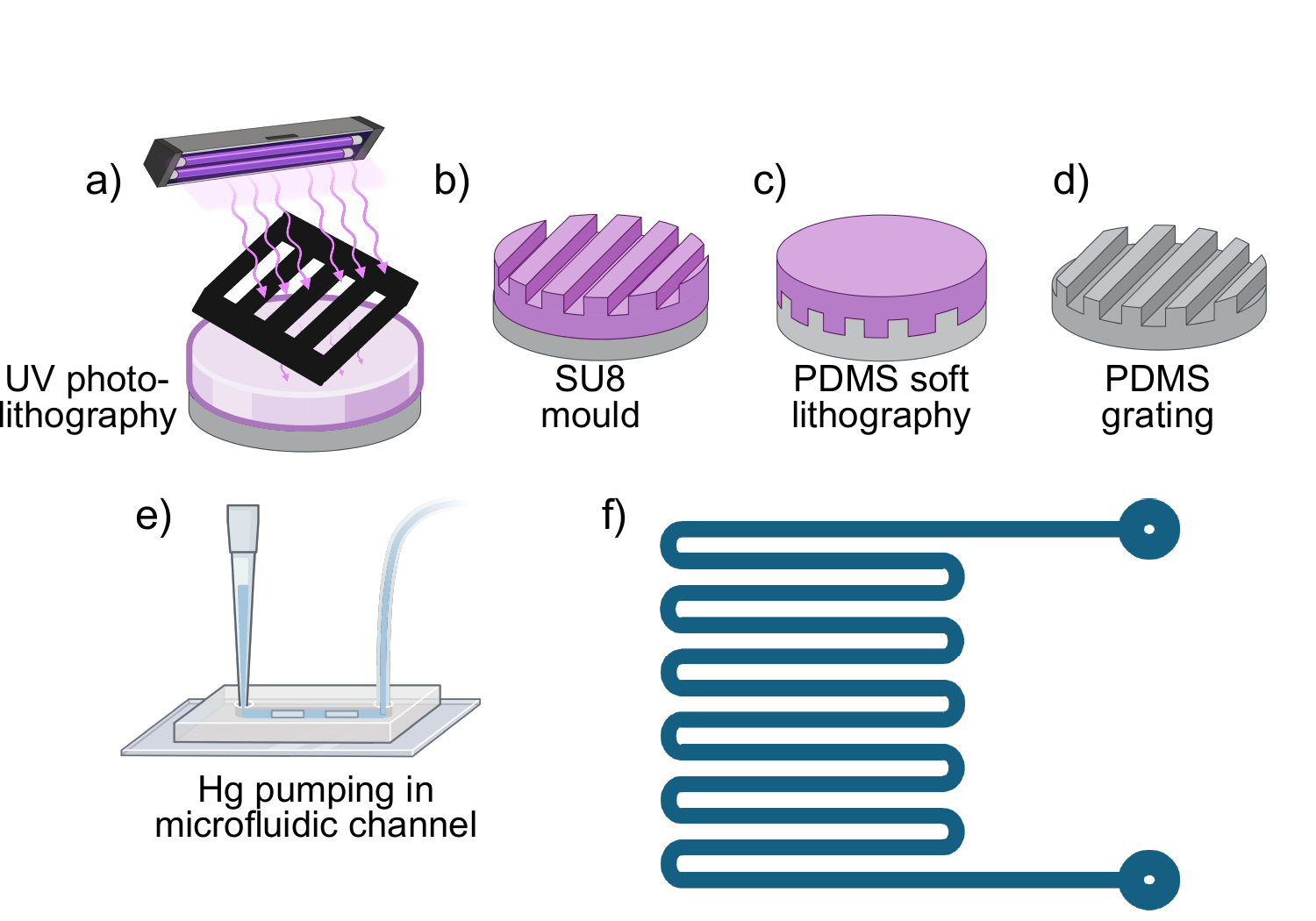}
    \caption{Fabrication process. \textbf{a} SU8 layer exposure to UV-photolithography using a photomask, resulting in a b) patterned SU8 mold. \textbf{c} Replication on PDMS via soft-lithography and (d) demolding. \textbf{e} Chip fabrication via bonding with glass and Hg pumping. \textbf{f} Microfluidic chip scheme of the design: a serpentine with two infusion/extraction channels to pump Hg in/out.}
    \label{fig:fabscheme}
\end{figure}

\subsection{Microfluidics channel}
The microfluidic device was designed with a periodicity of 79 $\upmu$m, with channel widths of 69 $\upmu$m and an inter-channel spacing of 10 $\upmu$m. These dimensions were chosen based on the prospective BT imaging system the devices would be tested in. Fabrication was carried out using standard UV and soft lithography processes. UV lithography was employed to produce a rigid polymeric mold, which served as a template for replication in PDMS (Figure \ref{fig:fabscheme}a-c). A two-dimensional CAD design containing the entire device layout was created using Rhinoceros CAD software and subsequently produced via photoplotting by Fine Line Imaging, Inc (Colorado, USA). The design consists of a continuous serpentine-shaped channel with a single inlet and outlet, connected to the serpentine via several-mm-long channels (see Figure \ref{fig:fabscheme}e). The serpentine design also enables eliminating transversal connections (often referred to as bridges) used as support in lithographic techniques, which can cause artifacts at the imaging stage \cite{peiffer2025applicability}. \\
An ordinary microscope glass slide was used as a substrate for the microfluidic chip. The glass was plasma treated and coated with a 25 $\upmu$m-thick layer of SU8-2050, a negative epoxy-based PR, via doctor blade coating. The coating was patterned using a mask aligner equipped with a 365 nm LED source at a dose of 160 mJ/cm$^2$ and the photomask produce by Fine Line Imaging, Inc. The coated substrate was baked before (soft baking) and after (post-exposure baking) exposure. Final development was carried out by immersing the sample in developer under gentle swirling for 6 min. The resulting SU8-mold was inspected using an optical microscope (Zeiss Axiolab 5 microscope) and a profilometer (Dektak 8, Veeco), and the results are shown in \textbf{Figure \ref{fig:characterization}a-b}, respectively. The former showed a periodicity of 79 $\upmu$m and spacings of 10 $\upmu$m as desired, and the profilometry across the channels detected a final thickness of $\sim$ 21 $\upmu$m. The thickness reduction from the nominal 25 $\upmu$m is attributed to solvent evaporation during processing \cite{mack2008fundamental}. \\
This mold was then used to fabricate a PDMS replica using Sylgard 184 at a 10:1 elastomer/curing agent ratio. The mixture was thoroughly mixed, degassed, poured onto the mold and thermally cured for 90 min. To facilitate the final demolding, the mold was previously treated with HMDS, a hydrophobic surfactant that minimizes the interaction with the PDMS-molded replica, by dip coating. After demolding, a microscope glass slide was placed on the PDMS-replica, and the assembly was exposed to plasma in air for 20 s to functionalize the surface and enhance adhesion. This resulted in a strong, covalent bond between the PDMS channels and the glass slide which ensured final encapsulation of Hg in the channels. The Hg was finally introduced in the channel using a peristaltic pump, as shown in Figure \ref{fig:characterization}c, where the Hg flowing inside the serpentine can be observed.  The three arrows in Figure \ref{fig:characterization}c highlight (from left to right) the serpentine-channel being filled, the loading channel connecting the serpentine and the inlet through which the Hg is enabled to flow from the peristaltic pump. \\

\subsection{Mask characterization}
The filled mask was characterized via optical microscopy and X-ray transmission imaging both in planar and CT scans. Images taken at the optical microscope reveal features in agreement with the ones found in the mold (Figure \ref{fig:characterization}d), proving an adequate mold replication and the feature preservation upon Hg filling. In both Figure \ref{fig:characterization}a and Figure \ref{fig:characterization}d, the curved regions of the serpentine are observed. While the optical image reveals some surface-level defects and minor damage, these are not expected to affect mask functionality and are in fact absent in the X-ray images, which confirms the presence of continuous, Hg-filled channels. Notably, both images were acquired over two years after fabrication, underscoring the long-term structural integrity of the device. This stability highlights the effectiveness of the PDMS-glass chip, which encapsulates the Hg and provides both mechanical robustness and a critical barrier for toxicity containment. \\
To confirm the complete filling of the channels with Hg a $\upmu$-CT scan was performed. This choice was preferred over other destructive methods such as electron microscopy to preserve mask integrity and avoid Hg spillage. The non-destructive $\upmu$-CT scan enabled us to fully resolve the channels across multiple millimeters through an effective pixel size of approximately 1.6 $\upmu$m; details on the used $\upmu$-CT system and data reconstruction are provided in the Experimental Section. The Hg channels were unsurprisingly characterized by a higher attenuation coefficient compared to the surrounding PDMS regions, as shown in Figure \ref{fig:characterization}e, allowing an easy discrimination between Hg and PDMS or glass regions, respectively highlighted in light purple and blue in the plot. Despite some variation across the channel profiles, the thickness of the Hg channels amounts to 21 $\upmu$m $\pm$ 2 $\upmu$m, matching well the SU-8 mold's depth shown in Figure \ref{fig:characterization}b. Two representative CT slices used to measure the Hg depth are presented in Supplementary Figure 1, and a 3D visualization of the corresponding mask region is available in Supplementary Video 1. This result proves (i) the full transfer of the serpentine features to the PMDS replica, and (ii) the complete filling of the channels. 
\begin{figure}
    \centering
    \includegraphics[width=.75\linewidth]{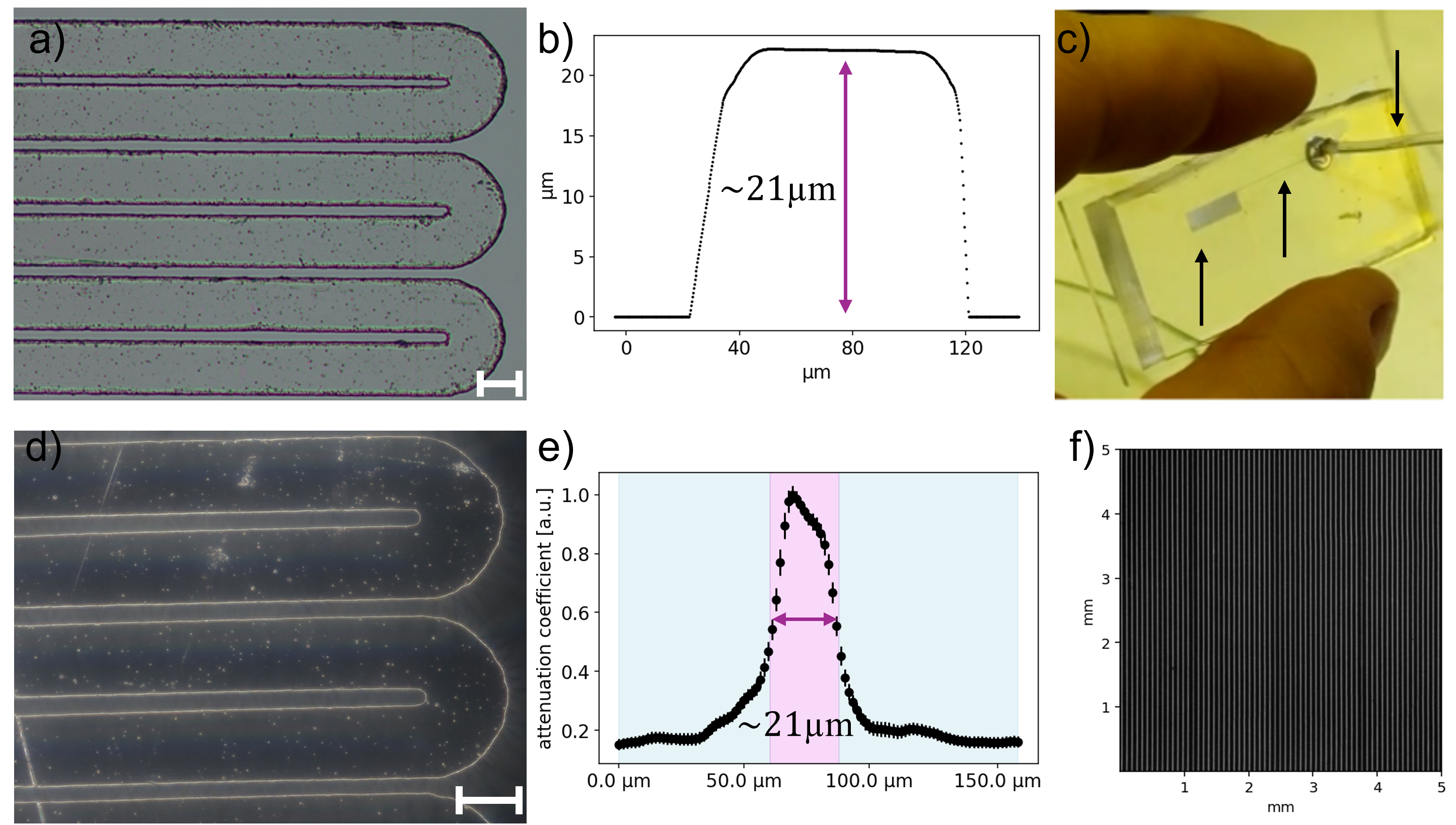}
    \caption{\textbf{a} Optical microscopy of the SU-8 mold. Scale-bar: 50 $\upmu$m. \textbf{b} Profilometry measurement of SU-8 mold inlet channel. The protrusive  $\sim$ 21 $\upmu$m channel is later imprinted in the PDMS chip. \textbf{c} Microfluidic chip during Hg pumping from the inlet to the serpentine through the loading channel (highlighted by black arrows). \textbf{d} Optical microscopy of the Hg-filled mask. Scale-bar: 50 $\upmu$m. \textbf{e} Attenuation coefficient profile of Hg channel (light purple shading) sandwiched between PDMS and glass (light blue shading), averaged over 1000 pixels and 10 channels. \textbf{f} Transmission-based radiography across multiple mm of the mask (pixel size = 5 $\upmu$m).}
    \label{fig:characterization}
\end{figure}
Finally, an X-ray transmission radiography was performed on the mask to investigate its performance with imaging systems at lower resolution. One of these images is shown in Figure \ref{fig:characterization}f, highlighting a good homogeneity of the channels over multiple millimeters. The mask's testing and employment in XPCI setups is reported in the next sections. 

\subsection{Beam Tracking imaging}
BT is a single-mask phase-based imaging approach, typically simpler to implement than the other grating-based techniques. The absorption grating selectively masks the X-ray beam creating individual beamlets whose sample-induced attenuation, refraction and scattering can be directly analyzed by a detector with pixel size smaller than the beamlet's width (\textbf{Figure \ref{fig:BT}a}, individual beamlets intensity shown in dark purple and pink for `reference' and `object' curves, respectively) by gaussian interpolation or statistical moments \cite{vittoria2015x}. These retrieval algorithms are discussed in more details in the Experimental Section. \\
\begin{figure}[h]
    \centering
    \includegraphics[width=.75\linewidth]{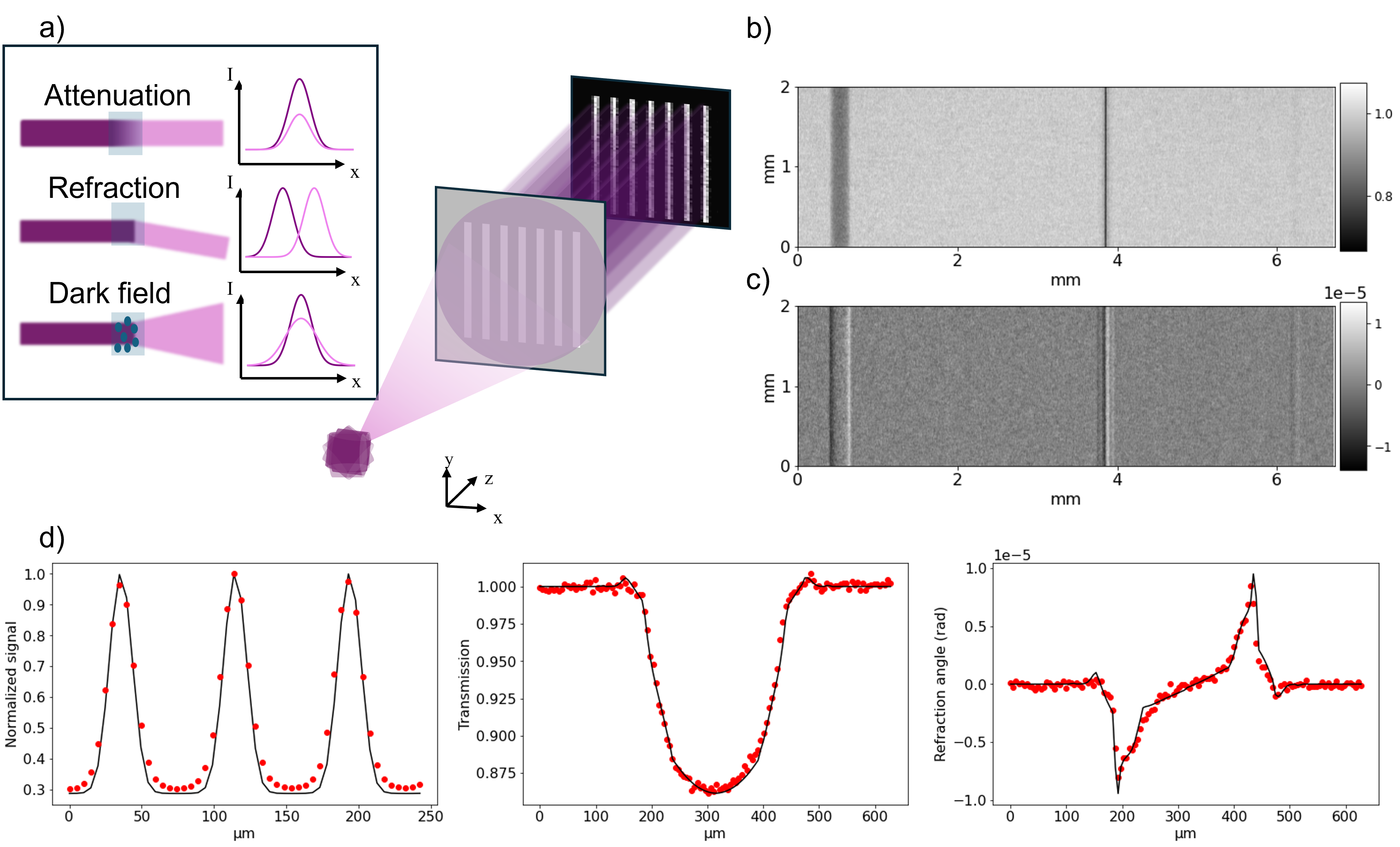}
    \caption{\textbf{a} Scheme of a BT imaging setup. The purple source emits a conical beam split by the grey mask into individual, well-separated beamlets resolved by the black detector at the back. In presence of a sample, the beamlets are perturbed allowing for analysis and retrieval. The sample's attenuation, refraction, and scattering effects on the beamlets are reported in the inset on the left. \textbf{b} The attenuation-based radiography of three wires i.e., (from left to right) sapphire, boron with a W-core, and nylon. \textbf{c} Corresponding refraction-based image, highlighting the edges and the nylon wire, almost invisible in the attenuation-based image. \textbf{d} Comparison between experimental data (red points) and simulated values (black profiles). The normalized profile on the left represent three unperturbed beamlets. The plots in the middle and the right show transmission and refraction through a 250 $\upmu$m wide sapphire wire.}
    \label{fig:BT}
\end{figure}
\indent Due to their simple geometry and homogeneous structure, cylindrical wires are often used as benchmark for phase-based imaging setups, as they allow for a straightforward comparison against simulated results \cite{diemoz2013sensitivity, kallon2015laboratory, wenz2015quantitative}. Three different sapphire, boron with a W-core, and nylon wires have been placed immediately downstream of the mask as shown in Figure \ref{fig:BT}a and imaged using a rotating anode X-ray source.  The wires (and all the X-ray images shown in this work) were imaged using a polychromatic 40 keV beam generated by a Mo anode with a mean delivered energy of approximately 21 kV. The retrieved attenuation and refraction images of the wires are shown in Figure \ref{fig:BT}b-c. The refraction contrast is consistently higher allowing to detect wires otherwise barely visible in attenuation, such as the thin nylon wire on the right in the two images. The different contrast channels were retrieved with a moment-based algorithm \cite{modregger2017interpretation}. The same system was then numerically simulated in a wave-optics based approach using the mask dimensions found during characterization. The same retrieval method was used to separate the different signals (transmission, refraction) for the simulated data and compared with the experimental ones, showing very good agreement as visible in Figure \ref{fig:BT}d. The figure presents, from left to right, the comparison of normalized intensity profiles across three mask periods unaffected by the sample, followed by the transmission and refraction profile of a sapphire wire, respectively; comparisons for the other wires can be found in Supplementary Figure 2. \\ 
\indent Maximizing the absorbing septa's stopping efficiency has been a persistent issue in X-ray imaging, as Au deposits via plating commonly suffer from sub-bulk densities resulting in lower visibility \cite{miao2014fabrication, michalska2023fabrication, buchanan2020effective}. This issue is worsened when alternative plating methods such as W powder centrifugal depositions are used and an even lower absorption efficiency is reached \cite{pinzek2022fabrication}. In this work we did not encounter any of these issues due to the Hg fluid physical state and bulk-like density maintained in the microfluidic channels, which enabled the maximum absorption given the aspect ratio obtained. Our interpretation is that the Hg fluid state minimizes incoherent scattering due to the absence of phase changes which are instead common in the granular structure of Au-plated and W-centrifugal deposits, thus making it the best candidate for X-ray selective absorption \cite{miao2014fabrication, pinzek2022fabrication, josell2020pushing, josell2021bottom}.  \\

\subsection{$\upmu$-CT scan}
The same BT set-up used for the wires was employed to perform $\upmu$-CT scans of two biological samples: a mouse lung (\textbf{Figure \ref{fig:CTscan}a-b}) and a piglet esophagus (Figure \ref{fig:CTscan}c-d). Samples were obtained from animals sacrificed for other, ethically approved studies, i.e., no animals were sacrificed specifically for this study. Indeed, images of the same specimens obtained with a different system can be found in Modregger et al.\cite{modregger2016small} and Savvidis et al.\cite{savvidis2022monitoring}, respectively. The same references also provide full details on the sample preparation methods. The samples were rotated over $360^{\circ}$ in 1200 equally spaced projections maintaining a 1s-long exposure for each projection, dithering the samples across a mask period in 8 and 16 steps for the lung and the esophagus, respectively. As can already be appreciated in the results in Figure \ref{fig:BT}, these masks can generate phase images in which the contrast is higher than in attenuation ones. In these CT scans, a multitude of details are made visible in the phase-based cross-sectional slices (Figure \ref{fig:CTscan}b,d) of both samples compared to their attenuation-based ones (Figure \ref{fig:CTscan}a,c). The phase-based image of the mouse lung highlights the bronchi, sharpening their edges and unveiling some left undetected by the attenuation-based counterpart. Similarly, the phase-based image of the piglet esophagus shows much sharper edges and a series of details completely undetected by the attenuation-based image in the epithelium and the submucosa region. The system's phase sensitivity can be further increased (i.e., leading to the same contrast values, but a higher contrast-to-noise ratio for the same exposure time \cite{kallon2020experimental}) by increasing the microfluidic grating's aspect ratio, which remains a goal for the future. \\
This image quality however provided a high enough contrast to render the imaged biological samples in 3D volumes using commercial softwares (see Experimental Section). Figure \ref{fig:CTscan}e shows a 3D volume render of the lung, where all the details in the airway tree are clearly resolved and displayed in a red-scale. Figure \ref{fig:CTscan}f displays the esophagus volume with a central cubic section exposed to reveal the internal submucosa structure, otherwise covered by the external walls, rendered in cyan.  
\end{justify} 
\begin{figure}
    \centering
    \includegraphics[width=\linewidth]{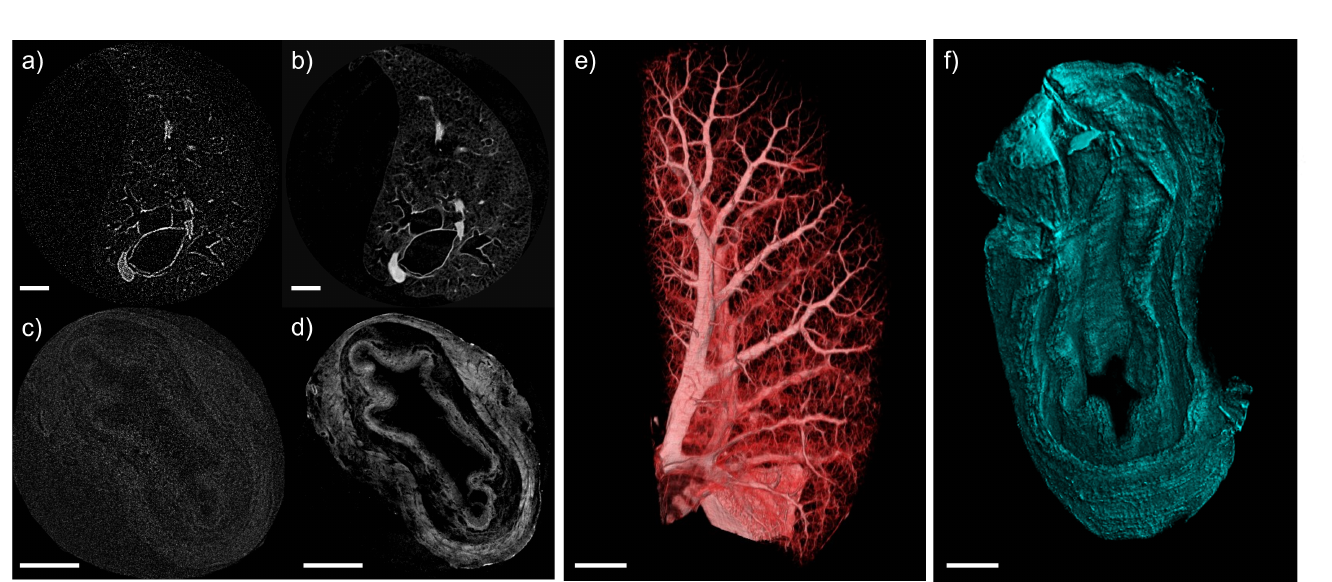}
    \caption{\textbf{a,b} Cross-sectional slices of the mouse lung obtained from attenuation- and phase-based reconstructions, respectively. \textbf{c,d} Corresponding attenuation- and phase-based slices of the piglet esophagus. Scale-bars: 2 mm. \textbf{e,f} Volumetric reconstruction of the lung and esophagus, respectively. Scale-bar: 1 mm.}
    \label{fig:CTscan}
\end{figure}

\section{Discussion}
\begin{justify}
This study demonstrates the feasibility and discusses the prospective advantages of employing microfluidic X-ray masks for XPCI. Their effectiveness for phase-based imaging was validated in a BT setup, enabling quantitative refraction-based image acquisition with enhanced sensitivity to low-Z samples (i.e., wires, mouse lung and esophagus) compared to the conventional attenuation-based counterpart. The deployment of these microfluidic masks is not limited to BT however, as several imaging modalities use absorption masks, thus promising a wide applicability across the field. \\  
\indent This fabrication method for microfluidic X-ray gratings is compatible with a variety of materials and processes, opening a yet untapped research path in XPCI. In the work described here, soft lithography with PDMS was used to replicate the lamellar grating structure, and Hg was introduced into the channels via peristaltic pumping to create the X-ray absorbing septa. However, alternative strategies can be employed to achieve similar grating functionality. Firstly, hard-PDMS has proven to increase the attainable AR by soft lithography compared to conventional PDMS used in this work \cite{lee2005pattern}. The lamellae could be also patterned directly in PR using standard UV or X-ray lithographic techniques, providing higher ARs. Epoxy-based PRs like SU8 can now commonly achieve few hundreds-micrometer high gratings \cite{joye2010uv} and also be used as molds for PDMS high-AR (AR $>$ 20) structures \cite{shao2012fabrication}. Other replication techniques include hot embossing and nanoimprint lithography, allowing direct transfer of high-AR features into thermoplastic materials, reaching sub-micrometer precision \cite{becker2000hot,chou1997sub, unno2023thermal}. \\
\indent Similarly, other high-Z contrast agents such as iodine-based solutions or metal alloys (e.g., Rose metal) may replace Hg to serve as the absorbing medium, particularly in situations where toxicity is a concern. This may lead to a reduction in visibility, but also e.g. exploit the increased attenuation at the K-edge energies of specific materials. It is important to emphasize that the amount of Hg required to fill meter-long microchannels is on the order of tens of milligrams, allowing for low exposure to Hg within most regulations (e.g., OSHA, NIOSH, EH40/2005 \cite{OSHA, EH40}). The pumping was performed under controlled ventilation to further mitigate the risk of vapor inhalation. Finally, the long-term integrity of the glass/PDMS chip developed in this work, which showed no leakage or damage over multiple years, demonstrates that Hg-filled microfluidic systems can be safely used without risk of inhalation or leakage, as long as sealing is maintained.\\
\indent While Hg peristaltic pumping offers reliable and controllable loading of the absorber medium, alternative approaches such as Hg porosimetry may be explored for applications requiring pressure-tuned filling. Such method would also allow an efficient fluid recovery from the microchannels and reuse, aligning with the principles of circular economy and safer toxicity management, requiring only a moderate applied pressure according to Young-Laplace law ($\sim$ 0.1 bar) \cite{washburn1921dynamics}. Additionally, the PDMS matrix (and protective layer) itself is recyclable under certain conditions \cite{feng2019multifunctional}, further supporting the sustainability of this fabrication approach, while ensuring year-long durability and a safe handling. The flexibility of this method, across both structural and absorbing materials, makes it adaptable to various imaging requirements and system constraints. Thus, the technique is not limited to the PDMS/Hg implementation described here but is broadly extensible to other material combinations and filling strategies.  \\
\indent More generally, as the interest in flexible X-ray imaging systems grows \cite{ciavatti2021high}, the need for adaptable and mechanically compliant optical components becomes increasingly important. Microfluidic gratings fabricated in elastomeric materials such as PDMS offer the compelling advantage of conforming to curved geometries without significant degradation. Unlike rigid gratings, which are limited by brittleness and can fracture under bending \cite{josell2024bottom}, these soft gratings can potentially withstand much larger radii of curvature, making them well-suited for integration into highly divergent X-ray imaging beams.
\end{justify}
\section{Conclusion}
\begin{justify}
In conclusion, this work demonstrates a simpler fabrication protocol for X-ray microfluidic gratings using soft lithography and peristaltic pumping. The proof-of-concept devices described in this paper unlocked higher contrast images phase-based imaging compared to conventional attenuation-based radiography despite the relatively low AR achieved. The translation to higher ARs will be the object of future work. We believe that this new approach will play a pivotal role in translating XPCI into industrial and clinical settings. The main reason lies in the ease of inexpensively scaling up this approach, which relies on a two-step process rather than the complex, multi-step workflows typically required. Currently, the cost of fabricating a few cm$^2$ of grating for XPCI remains in the range of the thousands of dollars, with even higher investments needed when developing or optimizing new processes. By contrast, our method significantly lowers both the entry barrier and per-unit cost (up to two orders of magnitude lower), allowing a broader adoption, rapid prototyping, and faster translation to practical imaging environments.


\section{Experimental Section}
\subsection{Fabrication process}
The microfluidic chip was fabricated via two lithographic-step processes: mold patterning via UV lithography and microfluidic channel replication via soft lithography. The rigid mold was fabricated in SU8-2050 coated on glass by doctor blade technique. The coated substrate was soft baked on a hot plate (Calctec, Italy) in two steps to reduce residual stress in the film. Firstly, it was baked at 65\textdegree C for 2 min and 30 s, followed by a temperature ramp of 10\textdegree C/min to 95\textdegree C and baking for 6 min. After cooling to room temperature, the SU8 was exposed by mask aligner (UV-KUB 2, Kloé, France) at 365 nm at with a dose of 160 mJ/cm using the serpentine-shaped photomask produced by Fine Line Imaging, Inc. The post-exposure bake was performed in two steps too on the same hotplate as the soft baking. It consisted of an initial baking at 65\textdegree C for 90 s followed by a 10\textdegree C/min ramp to 95\textdegree C and 6 min-long bake. Development was carried out by immersing the sample in SU8 developer under gentle manual agitation for 6 min. The PDMS-replica was prepared by mixing Sylgard 184 at a 10:1 elastomer/curing agent ratio, followed by degassing and curing at 70\textdegree C for 90 min. The SU8-mold was hydrophobically functionalized by dip-coating in a solution of HMDS dissolved in toluene (5\% molar concentration) for 5 s. 
The finalised microfluidic chip was filled with Hg by pumping using a peristaltic pump (Dolomite Mitos System, Unchained Labs, USA) connected to the loading channel in the chip via a Teflon tube. The outlet channel in the chip was also connected to a tube to accommodate overflow and collect possible Hg spilling.    
    \subsection{X-ray imaging}
The BT images (Figure \ref{fig:BT}b-d and \ref{fig:CTscan}) and the mask X-ray characterization (Figure \ref{fig:characterization}c) have been acquired on a customized imaging set-up using equipment available at the UCL Advanced X-Ray Imaging (AXIm) Labs following the scheme in Figure \ref{fig:BT}a. A Rigaku M007 X-ray source (Rigaku Corporation, Tokyo, Japan) with a rotating Mo target (70 $\upmu$m focal spot, 30 mA) provided a polychromatic beam at 40 kVp. A Hamamatsu (Hamamatsu, Shizuoka, Japan) C12849-111U detector with 10 $\upmu$m-thick GdO scintillator and 6.5x6.5 $\upmu$m$^2$ pixel size was used, enabling directly resolving the beamlets created by the pre-sample mask as required by the BT technique. Our mask was placed at \textit{d$_{1}$} = 0.54 m from the source and \textit{d}$_{2}$ = 0.17 m from the detector, amounting to a relatively compact set-up of total length of \textit{d}$_{tot}$ = \textit{d$_{1}$} + \textit{d$_{2}$} = 0.71 m. Such distances resulted in a mask magnification on the detector plane of \textit{M} = $\frac{d_{tot}}{d_{1}}$ $\approx$ 1.31. With the mask period being 79 $\upmu$m, the individual beamlets were well separated, with one mask period covering approximately 16 pixels on the detector plane. The samples were placed immediately downstream the mask to maximize the obtained refraction angles and scanned across one mask period in 8 (for the lung) and 16 (for the esophagus and the wires) dithering steps along the horizonal direction \textit{x} in Figure \ref{fig:BT}a. The sapphire, boron, and nylon wires (Goodfellows) are 250 $\upmu$m, 200 $\upmu$m, and 100 $\upmu$m in diameter, respectively. The W-core in the second wire has a 10 $\upmu$m diameter. \\
For the CT scans, the samples were rotated along the vertical axis (y-axis in Figure \ref{fig:BT}a) over 360° in 1200 evenly spaced projections. All the images were taken with 1s-long exposure times. \\
The mask X-ray $\upmu$-CT scan was performed using a Zeiss Xradia 620 Versa with a 4X magnification lens resulting in a (magnified) pixel size of approximately 1.6 $\upmu$m. The tube voltage was set at 160 kVp with a power of 17 W. Across the 2001 CT-projections taken, a High-Aspect-Ratio-Tomography (HART) strategy was adopted. HART is designed to increase the exposure time of a range of projections corresponding to the angles with the longest photon path through the sample. These projections, which in a planar sample like our mask correspond to the range of angles along which the mask's longer side is almost parallel to the beam, suffer from photon starvation resulting in artifacts in the image. Therefore, the HART strategy used here quadrupled the exposure time (20 s) on the appropriate range of angles, effectively suppressing these artifacts. The object reconstruction was performed using Zeiss proprietary software Advanced Reconstruction Toolbox.

    \subsection{Retrieval}
Two different retrieval methods were used in this work: the moment-based and the gaussian fitting methods \cite{modregger2016small, endrizzi2014hard}. The former method is based on the computation of the zeroth, first, and second statistical moment for each beamlet \textit{I} over its length \textit{l}, corresponding to approximately 16 pixels. These values represent the beamlet's area ($M_0$), center ($M_1$), and variance ($M_2$) as in the equations below:
\begin{subequations}
\begin{align}
M_0 &= \int_0^l I(x) \, dx \label{M0} \\
M_1 &= \frac{1}{M_0} \int_0^l x \cdot I(x) \, dx \label{M1}\\
M_2 &= \frac{1}{M_0} \int_0^l (x - M_1)^2 \cdot I(x) \, dx \label{M2}
\end{align}
\end{subequations}
The integrals are calculated over the same horizontal direction \textit{x} in Figure \ref{fig:BT}a. As the beamlet interact with the sample, the induced transmission (\textit{T}), refraction (\textit{R}), and scattering (\textit{S}) can be calculated by the following simple relations:
\begin{subequations}
\begin{align}
T &= M_0^{sample}/M_0 \label{Tmoment}\\
R &= M_1^{sample} - M_1 \label{Rmoment}\\
S &= M_2^{sample} - M_2 \label{Smoment}
\end{align}
\end{subequations}
where the $M_0^{sample}$, $M_1^{sample}$, and $M_2^{sample}$ are the moments calculated with the sample in the field of view. The images and data in Figure \ref{fig:BT}b-c were obtained using this method. \\
\indent The gaussian interpolation method is based instead on beamlet interpolation in the following gaussian equations:
\begin{subequations}
\begin{align}
    I(x) &= \frac{A}{\sqrt{2\pi \upsigma^2}} exp\left[-\frac{(x-x_0)^2}{2 \upsigma ^ 2}\right]  \label{Gauss_ref} \\
    I_{sample}(x) &= \frac{tA}{\sqrt{2\pi (\upsigma^2 + \upsigma_{sample}^2)}} exp\left[-\frac{(x -x_0 - \Delta x_{ref})^2}{2 (\upsigma ^ 2 + \upsigma_{sample}^2)}\right] \label{Gauss_obj}
\end{align}
\end{subequations}
with \ref{Gauss_ref} and \ref{Gauss_obj} interpolating the beamlets without and with sample, respectively. In these equations $\vec{x}$ is the gaussian's spatial extension vector and \textit{A}, $x_0$, and $\upsigma$ are its amplitude, center and width in absence of sample. When a sample is introduced, the curve is attenuated by the factor \textit{t}, shifted by $\Delta x_{ref}$, and its width is changed by $\upsigma_{sample}$. By fitting every beamlet to Equations \ref{Gauss_ref} and \ref{Gauss_obj} it is possible to find \textit{t}, $\Delta x_{ref}$, and $\upsigma_{sample}$ and to generate three independent images of the samples: attenuation, refraction, and scattering, respectively. \\
The CT scans were analyzed using the Gaussian fitting method on each individual projection. The projections were then reconstructed into transversal slices by inverse Radon transforms via python-supported scikit-image library. The resulting artifacts were blurred via selective blurring of the affected pixels using a Gaussian filter applied over five slices (approximately 25 $\upmu$m). In addition, residual artifacts were mostly removed y intensity-based thresholding, as their signal levels were consistently lower than those of relevant features. The slices were then rendered into 3D volumes using commercially available software: Avizo for the mouse lung and OSC Dragonfly for the esophagus. \\
The BT imaging system was simulated in Python using a wave-based approach based on the Fourier integral method \cite{vittoria2013strategies}, incorporating the previously described source and detector parameters. The mask dimensions used were those obtained from the characterization process. The energy spectrum delivered by the beam was taken from SpekPy 2.0.8 online toolkit inputting the source's characteristics. The mask, wire, and detector scintillator materials were taken from the Xraylib library. The simulation results, shown in Figure \ref{fig:BT}d, were compared with experimental data.

\medskip
\textbf{Supporting Information} \\
Supporting Information is available from the Wiley Online Library.

\medskip
\textbf{Acknowledgements} \\
This work was funded by the Nikon-UCL Prosperity Partnership on Next-Generation X-Ray Imaging, EPSRC (grant EP/T005408/1). The work was also supported by the European Research Council, ERC-StG-IntelGlazing grant no. 679891. A.O. was supported by the Royal Academy of Engineering under the “Chairs in Emerging Technologies” scheme (grant CiET1819/2/78). This work was also supported by the National Research Facility for Lab X-ray CT (NXCT) through EPSRC grant EP/T02593X/1. The authors also acknowledge the assistance of Dr. Carlos Navarrete-Leon in the 3D data visualization. Figure 1 and the Table of Content were assembled using BioRender templates.
\end{justify}

\medskip

\bibliographystyle{MSP}
\bibliography{refs}

\begin{thebibliography}{10}
\providecommand{\url}[1]{\texttt{#1}}
\providecommand{\urlprefix}{URL }

\bibitem{rontgen1896new}
W.~C. R{\"o}ntgen,
\newblock \emph{Science} \textbf{1896}, \emph{3}, 59 227.

\bibitem{bukreeva2016virtual}
I.~Bukreeva, A.~Mittone, A.~Bravin, G.~Festa, M.~Alessandrelli, P.~Coan, V.~Formoso, R.~G. Agostino, M.~Giocondo, F.~Ciuchi, et~al.,
\newblock \emph{Scientific reports} \textbf{2016}, \emph{6}, 1 27227.

\bibitem{villarraga2019x}
H.~Villarraga-G{\'o}mez, E.~L. Herazo, S.~T. Smith,
\newblock \emph{Precision Engineering} \textbf{2019}, \emph{60} 544.

\bibitem{lewis2004medical}
R.~A. Lewis,
\newblock \emph{Physics in medicine \& biology} \textbf{2004}, \emph{49}, 16 3573.

\bibitem{castelli2011mammography}
E.~Castelli, M.~Tonutti, F.~Arfelli, R.~Longo, E.~Quaia, L.~Rigon, D.~Sanabor, F.~Zanconati, D.~Dreossi, A.~Abrami, et~al.,
\newblock \emph{Radiology} \textbf{2011}, \emph{259}, 3 684.

\bibitem{willer2021x}
K.~Willer, A.~A. Fingerle, W.~Noichl, F.~De~Marco, M.~Frank, T.~Urban, R.~Schick, A.~Gustschin, B.~Gleich, J.~Herzen, et~al.,
\newblock \emph{The Lancet Digital Health} \textbf{2021}, \emph{3}, 11 e733.

\bibitem{shoukroun2020enhanced}
D.~Shoukroun, L.~Massimi, F.~Iacoviello, M.~Endrizzi, D.~Bate, A.~Olivo, P.~Fromme,
\newblock \emph{Composites Part B: Engineering} \textbf{2020}, \emph{181} 107579.

\bibitem{partridge2022enhanced}
T.~Partridge, A.~Astolfo, S.~Shankar, F.~Vittoria, M.~Endrizzi, S.~Arridge, T.~Riley-Smith, I.~Haig, D.~Bate, A.~Olivo,
\newblock \emph{Nature communications} \textbf{2022}, \emph{13}, 1 4651.

\bibitem{david2007fabrication}
C.~David, J.~Bruder, T.~Rohbeck, C.~Gr{\"u}nzweig, C.~Kottler, A.~Diaz, O.~Bunk, F.~Pfeiffer,
\newblock \emph{Microelectronic Engineering} \textbf{2007}, \emph{84}, 5-8 1172.

\bibitem{pfeiffer2006phase}
F.~Pfeiffer, T.~Weitkamp, O.~Bunk, C.~David,
\newblock \emph{Nature physics} \textbf{2006}, \emph{2}, 4 258.

\bibitem{miao2014fabrication}
H.~Miao, A.~A. Gomella, N.~Chedid, L.~Chen, H.~Wen,
\newblock \emph{Nano letters} \textbf{2014}, \emph{14}, 6 3453.

\bibitem{michalska2023fabrication}
M.~Michalska, A.~Rossi, G.~Kokot, C.~M. Macdonald, S.~Cipiccia, P.~R. Munro, A.~Olivo, I.~Papakonstantinou,
\newblock \emph{Advanced functional materials} \textbf{2023}, \emph{33}, 16 2212660.

\bibitem{josell2021bottom}
D.~Josell, Z.~Shi, K.~Jefimovs, V.~A. Guzenko, C.~Beauchamp, L.~Peer, M.~Polikarpov, T.~P. Moffat,
\newblock \emph{Journal of the Electrochemical Society} \textbf{2021}, \emph{168}, 8 082508.

\bibitem{rossi2025fabrication}
A.~Rossi, I.~Buchanan, A.~Astolfo, M.~Michalska, D.~Briglin, A.~Charman, D.~Josell, A.~Olivo, I.~Papakonstantinou,
\newblock \emph{Advanced Materials Interfaces} \textbf{2025}, \emph{12}, 8 2400749.

\bibitem{pereira2025quantifying}
A.~Pereira, S.~Spindler, Z.~Shi, L.~Romano, M.~Rawlik, F.~Marone, D.~Josell, M.~Stauber, M.~Stampanoni,
\newblock \emph{Scientific Reports} \textbf{2025}, \emph{15}, 1 14223.

\bibitem{noda2008fabrication}
D.~Noda, M.~Tanaka, K.~Shimada, W.~Yashiro, A.~Momose, T.~Hattori,
\newblock \emph{Microsystem technologies} \textbf{2008}, \emph{14} 1311.

\bibitem{josell2020pushing}
D.~Josell, Z.~Shi, K.~Jefimovs, L.~Romano, J.~Vila-Comamala, T.~P. Moffat,
\newblock \emph{Journal of The Electrochemical Society} \textbf{2020}, \emph{167}, 13 132504.

\bibitem{romano2020high}
L.~Romano, J.~Vila-Comamala, K.~Jefimovs, M.~Stampanoni,
\newblock \emph{Advanced Engineering Materials} \textbf{2020}, \emph{22}, 10 2000258.

\bibitem{laermer2020deep}
F.~Laermer, S.~Franssila, L.~Sainiemi, K.~Kolari,
\newblock In \emph{Handbook of silicon based MEMS materials and technologies}, 417--446. Elsevier, \textbf{2020}.

\bibitem{li2000metal}
X.~Li, P.~Bohn,
\newblock \emph{Applied Physics Letters} \textbf{2000}, \emph{77}, 16 2572.

\bibitem{huang2011metal}
Z.~Huang, N.~Geyer, P.~Werner, J.~De~Boor, U.~G{\"o}sele,
\newblock \emph{Advanced materials} \textbf{2011}, \emph{23}, 2 285.

\bibitem{romankiw1997plating}
L.~T. Romankiw, E.~J. O’Sullivan,
\newblock \emph{Handbook of Microlithography, Micromachining, and Microfabrication} \textbf{1997}, \emph{2} 197.

\bibitem{vila2018towards}
J.~Vila-Comamala, L.~Romano, V.~Guzenko, M.~Kagias, M.~Stampanoni, K.~Jefimovs,
\newblock \emph{Microelectronic Engineering} \textbf{2018}, \emph{192} 19.

\bibitem{kashyap2016experimental}
Y.~Kashyap, H.~Wang, K.~Sawhney,
\newblock \emph{Optics express} \textbf{2016}, \emph{24}, 16 18664.

\bibitem{wang2016synchrotron}
H.~Wang, Y.~Kashyap, K.~Sawhney,
\newblock \emph{Scientific reports} \textbf{2016}, \emph{6}, 1 20476.

\bibitem{de2023facile}
P.~de~Haan, K.~Mathwig, L.~Yuan, B.~W. Peterson, E.~Verpoorte,
\newblock \emph{Organs-on-a-Chip} \textbf{2023}, \emph{5} 100026.

\bibitem{xia1998soft}
Y.~Xia, G.~M. Whitesides,
\newblock \emph{Angewandte Chemie International Edition} \textbf{1998}, \emph{37}, 5 550.

\bibitem{josell2024bottom}
D.~Josell, D.~Raciti, T.~Gnaupel-Herold, A.~Pereira, V.~Tsai, Q.~Yu, L.~Chen, M.~Stauber, M.~Rawlik, M.~Stampanoni, et~al.,
\newblock \emph{Journal of The Electrochemical Society} \textbf{2024}, \emph{171}, 3 032502.

\bibitem{schroter2017large}
T.~J. Schr{\"o}ter, F.~J. Koch, P.~Meyer, D.~Kunka, J.~Meiser, K.~Willer, L.~Gromann, F.~De~Marco, J.~Herzen, P.~Noel, et~al.,
\newblock \emph{Review of Scientific Instruments} \textbf{2017}, \emph{88}, 1.

\bibitem{united2010unscear}
U.~N. S.~C. on~the Effects~of Atomic~Radiation, et~al.,
\newblock \emph{Report to the General Assembly} \textbf{2010}, \emph{1}.

\bibitem{peiffer2025applicability}
C.~Peiffer, A.~Astolfo, M.~Endrizzi, C.~Hagen, A.~Olivo, P.~Munro,
\newblock \emph{Journal of Physics D: Applied Physics} \textbf{2025}, \emph{58}, 31 315402.

\bibitem{mack2008fundamental}
C.~Mack,
\newblock \emph{Fundamental principles of optical lithography: the science of microfabrication},
\newblock John Wiley \& Sons, \textbf{2008}.

\bibitem{vittoria2015x}
F.~A. Vittoria, M.~Endrizzi, P.~C. Diemoz, A.~Zamir, U.~H. Wagner, C.~Rau, I.~K. Robinson, A.~Olivo,
\newblock \emph{Scientific reports} \textbf{2015}, \emph{5}, 1 16318.

\bibitem{diemoz2013sensitivity}
P.~Diemoz, C.~Hagen, M.~Endrizzi, A.~Olivo,
\newblock \emph{Applied Physics Letters} \textbf{2013}, \emph{103}, 24.

\bibitem{kallon2015laboratory}
G.~K. Kallon, M.~Wesolowski, F.~A. Vittoria, M.~Endrizzi, D.~Basta, T.~P. Millard, P.~C. Diemoz, A.~Olivo,
\newblock \emph{Applied Physics Letters} \textbf{2015}, \emph{107}, 20.

\bibitem{wenz2015quantitative}
J.~Wenz, S.~Schleede, K.~Khrennikov, M.~Bech, P.~Thibault, M.~Heigoldt, F.~Pfeiffer, S.~Karsch,
\newblock \emph{Nature communications} \textbf{2015}, \emph{6}, 1 7568.

\bibitem{modregger2017interpretation}
P.~Modregger, M.~Kagias, S.~C. Irvine, R.~Br{\"o}nnimann, K.~Jefimovs, M.~Endrizzi, A.~Olivo,
\newblock \emph{Physical review letters} \textbf{2017}, \emph{118}, 26 265501.

\bibitem{buchanan2020effective}
I.~Buchanan, G.~Kallon, T.~Beckenbach, J.~Schulz, M.~Endrizzi, A.~Olivo,
\newblock \emph{Journal of Applied Physics} \textbf{2020}, \emph{128}, 21.

\bibitem{pinzek2022fabrication}
S.~Pinzek, A.~Gustschin, N.~Gustschin, M.~Viermetz, F.~Pfeiffer,
\newblock \emph{Scientific Reports} \textbf{2022}, \emph{12}, 1 5405.

\bibitem{modregger2016small}
P.~Modregger, T.~P. Cremona, C.~Benarafa, J.~C. Schittny, A.~Olivo, M.~Endrizzi,
\newblock \emph{Scientific reports} \textbf{2016}, \emph{6}, 1 30940.

\bibitem{savvidis2022monitoring}
S.~Savvidis, M.~F. Gerli, M.~Pellegrini, L.~Massimi, C.~K. Hagen, M.~Endrizzi, A.~Atzeni, O.~K. Ogunbiyi, M.~Turmaine, E.~S. Smith, et~al.,
\newblock \emph{Acta Biomaterialia} \textbf{2022}, \emph{141} 290.

\bibitem{kallon2020experimental}
G.~Kallon, F.~Vittoria, I.~Buchanan, M.~Endrizzi, A.~Olivo,
\newblock \emph{Journal of Physics D: Applied Physics} \textbf{2020}, \emph{53}, 19 195404.

\bibitem{lee2005pattern}
T.-W. Lee, O.~Mitrofanov, J.~W. Hsu,
\newblock \emph{Advanced Functional Materials} \textbf{2005}, \emph{15}, 10 1683.

\bibitem{joye2010uv}
C.~D. Joye, J.~P. Calame, M.~Garven, B.~Levush,
\newblock \emph{Journal of Micromechanics and Microengineering} \textbf{2010}, \emph{20}, 12 125016.

\bibitem{shao2012fabrication}
G.~Shao, J.~Wu, Z.~Cai, W.~Wang,
\newblock \emph{Sensors and Actuators A: Physical} \textbf{2012}, \emph{178} 230.

\bibitem{becker2000hot}
H.~Becker, U.~Heim,
\newblock \emph{Sensors and Actuators A: Physical} \textbf{2000}, \emph{83}, 1-3 130.

\bibitem{chou1997sub}
S.~Y. Chou, P.~R. Krauss, W.~Zhang, L.~Guo, L.~Zhuang,
\newblock \emph{Journal of Vacuum Science \& Technology B: Microelectronics and Nanometer Structures Processing, Measurement, and Phenomena} \textbf{1997}, \emph{15}, 6 2897.

\bibitem{unno2023thermal}
N.~Unno, T.~M{\"a}kel{\"a},
\newblock \emph{Nanomaterials} \textbf{2023}, \emph{13}, 14 2031.

\bibitem{OSHA}
O.~Safety, H.~Administration,
\newblock Occupational chemical database,
\newblock \urlprefix\url{{https://www.osha.gov/chemicaldata/505}},
\newblock Accessed: 2025-07-07.

\bibitem{EH40}
Health, S.~Executive,
\newblock Eh40/2005 workplace exposure limits, \textbf{2020},
\newblock \urlprefix\url{{https://www.hse.gov.uk/pubns/priced/eh40.pdf}},
\newblock Accessed: 2025-07-07.

\bibitem{washburn1921dynamics}
E.~W. Washburn,
\newblock \emph{Physical review} \textbf{1921}, \emph{17}, 3 273.

\bibitem{feng2019multifunctional}
Z.~Feng, B.~Yu, J.~Hu, H.~Zuo, J.~Li, H.~Sun, N.~Ning, M.~Tian, L.~Zhang,
\newblock \emph{Industrial \& Engineering Chemistry Research} \textbf{2019}, \emph{58}, 3 1212.

\bibitem{ciavatti2021high}
A.~Ciavatti, R.~Sorrentino, L.~Basiric{\`o}, B.~Passarella, M.~Caironi, A.~Petrozza, B.~Fraboni,
\newblock \emph{Advanced Functional Materials} \textbf{2021}, \emph{31}, 11 2009072.

\bibitem{endrizzi2014hard}
M.~Endrizzi, P.~C. Diemoz, T.~P. Millard, J.~Louise~Jones, R.~D. Speller, I.~K. Robinson, A.~Olivo,
\newblock \emph{Applied Physics Letters} \textbf{2014}, \emph{104}, 2.

\bibitem{vittoria2013strategies}
F.~A. Vittoria, P.~C. Diemoz, M.~Endrizzi, L.~Rigon, F.~C. Lopez, D.~Dreossi, P.~R. Munro, A.~Olivo,
\newblock \emph{Applied optics} \textbf{2013}, \emph{52}, 28 6940.

\end{thebibliography}

\end{document}